\documentclass{aa}

\usepackage{txfonts}
\usepackage{graphicx}

\begin{document}
\def\gapprox{\;\rlap{\lower 2.5pt            
    \hbox{$\sim$}}\raise 1.5pt\hbox{$>$}\;}       
\def\lapprox{\;\rlap{\lower 2.5pt            
    \hbox{$\sim$}}\raise 1.5pt\hbox{$<$}\;}

\title{Energetic radiation and the sulfur chemistry of
protostellar envelopes: Submillimeter interferometry of AFGL 2591}
\author{A. O. Benz\inst{1}, P. St\"auber \inst{1}, T. L. Bourke\inst{2}, F. F. S. van der Tak\inst{3,4}, E. F. van Dishoeck\inst{5} and J. K. J{\o}rgensen\inst{2}}
\offprints{A.O. Benz, \email{benz@astro. phys.ethz.ch}}
\institute{Institute of Astronomy, ETH, CH-8092 Zurich, Switzerland 
\and Harvard-Smithsonian Center for Astrophysics, 60 Garden Street, Cambridge, MA 02138, USA
\and Max-Planck Institut f\"ur Radioastronomie, auf dem H\"ugel 69, 53121 Bonn, Germany
\and Netherlands Institute for Space Research (SRON), Landleven 12, 9747 AD Groningen, The Netherlands
\and Leiden Observatory, Leiden University, PO Box 9513, NL-2300 RA Leiden, The Netherlands
}

\date{Received / Accepted }

\abstract{The chemistry in the inner few thousand AU of accreting envelopes around young stellar objects is predicted to vary greatly with far-UV and X-ray irradiation by the central star.}{We search for molecular tracers of high-energy irradiation by the protostar in the hot inner envelope.}{The Submillimeter Array (SMA) has observed the high-mass star forming region AFGL 2591 in lines of CS, SO, HCN, HCN($\nu_2$=1), and HC$^{15}$N with 0.6$''$ resolution at 350 GHz probing radial scales of 600-3500 AU for an assumed distance of 1 kpc. The SMA observations are compared with the predictions of a chemical model fitted to previous single-dish observations.}{The CS and SO main peaks are extended in space at the FWHM level, as predicted in the model assuming protostellar X-rays. However, the main peak sizes are found smaller than modeled by nearly a factor of 2. On the other hand, the lines of CS, HCN, and HC$^{15}$N, but not SO and HCN($\nu_2$=1), show pedestal emissions at radii $\lapprox$ 3500 AU that are not predicted.  All lines except SO show a secondary peak within the approaching outflow cone. A dip or null in the visibilities caused by a sharp decrease in abundance with increasing radius is not observed in CS and only tentatively in SO.}{The emission of protostellar X-rays is supported by the good fit of the modeled SO and CS amplitude visibilities including an extended main peak in CS. The broad pedestals can be interpreted by far-UV irradiation in a spherically non-symmetric geometry, possibly comprising outflow walls on scales of 3500 -- 7000 AU. The extended CS and SO main peaks suggest sulfur evaporation near the 100 K temperature radius. The effects of the corresponding abundance jumps may be reduced in visibility plots by smoothing due to inhomogeneity at the evaporation radius, varying by $\pm$10\% or more in different directions. 
}
\titlerunning{SMA observations of AFGL 2591}
\authorrunning{A. O. Benz et al.}
\maketitle

\section{Introduction}

X-ray emission has been reported from high-mass objects of the type UCHII and Herbig Ae/Be, representing the late phases of star formation, with X-ray luminosities in the range $10^{30} - 10^{33}$ erg s$^{-1}$ (Yamauchi \& Koyama 1993; Carkner, Kozak \& Feigelson 1998; Hofner et al. 2002). Possible origins of the X-rays include magnetic activity due to star-disk interaction, accretions shocks, and shocks in outflows. The high mass makes it conceivable that the central object is also an emitter of substantial UV radiation. The inner envelope may thus experience high-energy irradiation within a few 1000 AU from the central object, where processes important in the formation of a star take place: molecules evaporate, a disk may form, and outflows and winds are accelerated. With subarcsecond spatial resolution observations using (sub)millimeter wavelength interferometers at high-altitude sites, it has become possible to disentangle these phenomena. 

Younger high-mass protostellar objects are particularly interesting in terms of their chemistry: before the central star is formed most molecules are frozen out on dust grains but evaporate when the young star ignites and starts heating its ambient envelope. As a result of this, objects in this "Hot Core" phase show a complex spectrum of molecular line emission at (sub)millimeter wavelengths (van Dishoeck 2003). High-Mass Protostellar Objects (HMPOs) are bright in mid-infrared (10-12 $\mu$m) and show a central density concentration. An open question remains how important UV photons and X-rays are for regulating the chemistry in these earliest phases. In the Hot Core and HMPO phase, a protostellar object is still deeply embedded in a collapsing dusty envelope. Thus UV photons and X-rays are not able to escape and are not directly observable. Their irradiation, however, may change severely the chemistry of the envelope (Maloney et al. 1996). 

Here we focus on AFGL 2591, classified as an HMPO or early Hot Core. It is a well studied object and a testbed for high-mass star formation. AFGL 2591 is located in a relatively isolated region at a distance of 0.5 to 2 kpc (as discussed by van der Tak et al. 1999). A recent estimate puts it at 1.7 kpc (Schneider et al. 2006). As this is not confirmed, we will follow the older literature and use 1 kpc in the following for convenience. Numerous molecular species have been observed in AFGL 2591 (for a summary see Doty et al. 2002). HCN observed by infrared absorption and by submillimeter emission lines reaches a high abundance ($\approx 10^{-6}$) in the central core, having a radius less than 175 AU and a temperature of more than 300 K (Lahuis \& van Dishoeck 2000; Boonman et al. 2001; Knez et al. 2003). On the other hand, CS 5-4 emission has been reported to have a half-power radius of 26 000 AU in single-dish mapping (Carr et al. 1995). The two extremes indicate the approximate extent of the molecular envelope. 

Thermal desorption not only changes molecular abundances in the gas phase. It makes these molecules vulnerable to energetic radiation, converting the predominantly neutral Hot Core chemistry into ion-chemistry. Recently, St\"auber et al. (2004a, 2005) have fitted chemical models for AFGL 2591 to column densities derived from single-dish line fluxes of 24 molecules and included the effects of energetic radiation. The fits improve significantly under the assumption of protostellar far-ultraviolet (6 -- 13.6 eV) emission and substantial X-ray luminosities ($6\cdot 10^{31} - 10^{32}$ erg s$^{-1}$). The spherically symmetric chemical model of St\"auber et al. (2004a) predicts that far-UV emission reaches as far as about 350 AU from the central object. The discovery of CO$^+$ and SO$^+$ emission from AFGL 2591 (St\"auber et al. 2004b, 2007) is further evidence for the presence of protostellar far-UV emission. X-rays can penetrate farther into the envelope than far-UV since the cross section strongly decreases with energy up to 20 keV (Morrison \& McCammon 1983). Protostellar X-rays have noticeable effects in the envelope out to about 5000 AU, from where on ionization by cosmic rays dominates. Thus the ionization rate changes significantly over the range of a few hundred AU to a few thousand AU, leaving unique signatures in the spatial profile of individual molecular species. 

According to the chemical models of St\"auber et al. (2005), CS and SO are sensitive to reactions with water plus energetic radiation: In the region where $T > 100$ K, CS and SO are reduced by water in gas phase. However, if this region is irradiated by X-rays from the central object, the abundances of CS and SO increase greatly. The sulfur-bearing molecules are then enhanced by reactions of atomic sulfur -- produced by X-ray induced far-UV photodissociation -- reacting with OH, and the product SO reacting with atomic carbon. In a beam with 1$''$ (1000 AU) radius, the model abundances of CS and SO are enhanced by more than an order of magnitude, if irradiated by a central X-ray luminosity of some $10^{32}$ erg s$^{-1}$. The enhancement by protostellar far-UV emission is at most a factor of a few for CS and practically none for SO. For HCN the result is inverse. Thus, SO is a good tracer for X-rays, HCN is a tracer for far-UV. CS is greatly enhanced by X-rays, but effectively traces both high-energy radiations. 

We present new observations of AFGL 2591 with 0.6$''$ ($\approx$ 600 Astronomical Units, AU) spatial resolution. The goal of these investigations is to confront the predictions of the model of St\"auber et al. (2005) reproducing single-dish observations with new interferometric measurements that actually probe the relevant scales. To avoid confusion with low density material plaguing low-$J$ data, observations in submillimeter waves are necessary. Observational details are given in Section 2. The continuum observations, briefly summarized in Section 3, are used as a reference for the peak location of the molecular lines. The observed spatial profiles of molecular species, presented as visibility amplitudes in Section 4, can then be used as a diagnostic for high-energy radiations from the YSO. In Section 5, chemical models are presented and the relation between geometry and visibility amplitudes is studied. Observations and chemical models are compared in Section 6 and the results are discussed in Section 7. Conclusions on the capabilities of chemical modeling under high-energy irradiation concerning sulfur chemistry are drawn in Section 8. This paper reports first results and will serve as a reference for the analysis of future SMA observations and modeling of AFGL 2591.

\begin{figure*}[t]
\centering
\resizebox{16cm}{!}{\includegraphics[angle=270]{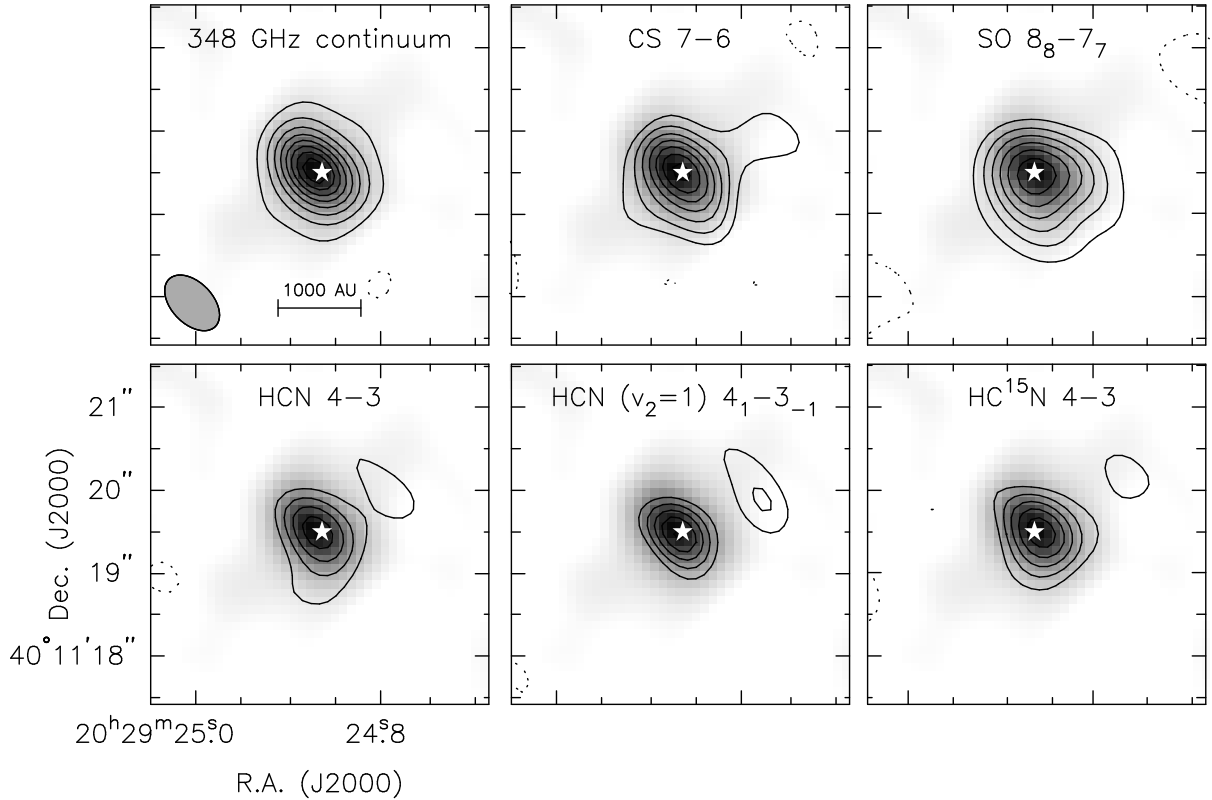}}
\caption{Clean maps of AFGL 2591 emissions observed with the SMA. All images display the 348 GHz continuum in grayscale for reference. The contours show either the continuum or line emission as indicated in each panel. Contours for the continuum are 3$\sigma$, 6$\sigma$, 9$\sigma$, etc. For the line data they are 3$\sigma$, 5$\sigma$, 7$\sigma$, etc. Negative isophotes (-3$\sigma$) are dashed. The rms noise in the line maps is on the average 0.12 Jy/beam. The white asterisk indicates the position of VLA 3 measured at 205 GHz by van der Tak et al.\ (2006). The half-power beam is given in the first image, as well as the length scale assuming a 1 kpc distance.} 
\end{figure*}

\section{Observations}

The Submillimeter Array (SMA)\footnote{The Submillimeter Array is a joint project between the Smithsonian Astrophysical Observatory and the Academia Sinica Institute of Astronomy and Astrophysics and is funded by the Smithsonian Institution and the Academia Sinica.} (Ho et al. 2004) on the top of Mauna Kea in Hawaii observed AFGL 2591 in the frequency bands of 342.6 - 344.6 GHz and 352.6 - 354.6 GHz on May 13, 2005. The frequency range includes high-$J$ lines of the sulfur-bearing molecules SO and CS, as well as two lines of HCN and one of HC$^{15}$N (Tab. 1). These lines, sensitive to X-rays or far-UV irradiation, allow us to identify protostellar irradiation and to separate the effects of X-ray and far-UV emissions. CO$^+$ is also in the range, but was not clearly detected. 

The primary frequency resolution was 0.406 MHz (0.5 km s$^{-1}$) and uniform across the 2 GHz band. 3C279 was used for flux calibration, 3C279 and Jupiter for bandpass calibration. Gain calibration was performed using the quasars BL Lac and J2015+371. The flux calibration uncertainty is of the order of 10\%, but could be occasionally as high as 20\%. A total of 5 elements of the interferometer were used in the extended array configuration covering projected baselines from 38 k$\lambda$ to 215 k$\lambda$ (32.7 m to 185 m). The FWHM beam size for natural weighting is 0.8$''\times\ 0.5''$ with a position angle of 44 degrees. The field of view (FWHM beam of single antenna) is 39$''$. The resolution in extended array is ideal for the Hot Core envelope from a few hundred to a few thousand AU; the interferometric resolution allows to probe spatial scales in the emission down to about 600 AU in radius, and it will miss structures having radii much larger than 3500 AU due to missing short baselines. The overall telescope time was 5.6 hours in fair weather at PWV $\approx$ 2 mm. The rms noise in the continuum map is 16 mJy/beam. As there is no (single dish) total power mode available for the SMA, we use previously reported values from the James Clerk Maxwell Telescope (JCMT) to estimate the missing flux due to the lack of short spacings.

The data were reduced using the SMA MIR package (Qi 2005). For image reconstruction the MIRIAD software\footnote{http://sma-www.cfa.harvard.edu/miriadWWW/manuals/ SMAuguide/smauserhtml/} (Sault et al. 1995) was applied. The continuum emission was sufficiently strong to allow for self-calibration. It was subsequently used to improve also the image quality of line observations.

\section{Continuum observations}
The observations of the continuum are briefly presented. They are used here as the reference for the location of the line emission (Fig. 1) and for self-calibration. The results will be discussed in a later paper. Figure 1a displays the map of the continuum emission after combining both sidebands. A Gaussian fit to the emission in the $(u,v)$ plane yields a total flux density of 0.662$\pm0.013$ Jy, where the error refers to the statistics and does not include calibration. The peak flux density in the image (Fig. 1a) is 0.30$\pm0.02$ Jy per synthesized beam, identical for the uncleaned and cleaned images in natural and uniform weighting. The peak is located at RA(J2000) = 20$^h$29$^m$24.$^s$880(7) and Dec(J2000) = +40$^\circ 11'19.''5(1)$ as fitted in $(u,v)$ data. The values given in parenthesis are the mean error of the last digit resulting from calibration errors. The position corresponds to peak VLA3 by Trinidad et al. (2003) and is also consistent with the measurements at 205 GHz by van der Tak et al. (2006). As the peak is not in the phase center, an offset was applied in all subsequent procedures in $(u,v)$ space. The peak value of the continuum emission amounts to a brightness temperature of 10.4 K, using Planck's law. The integrated continuum flux at 348 GHz is consistent with thermal dust emission and follows the increase with frequency reported before in millimeter waves (van der Tak et al. 1999).

\begin{figure}[t]
\centering
\begin{minipage}[t]{0.50\linewidth}
\centering
\includegraphics[width=\linewidth]{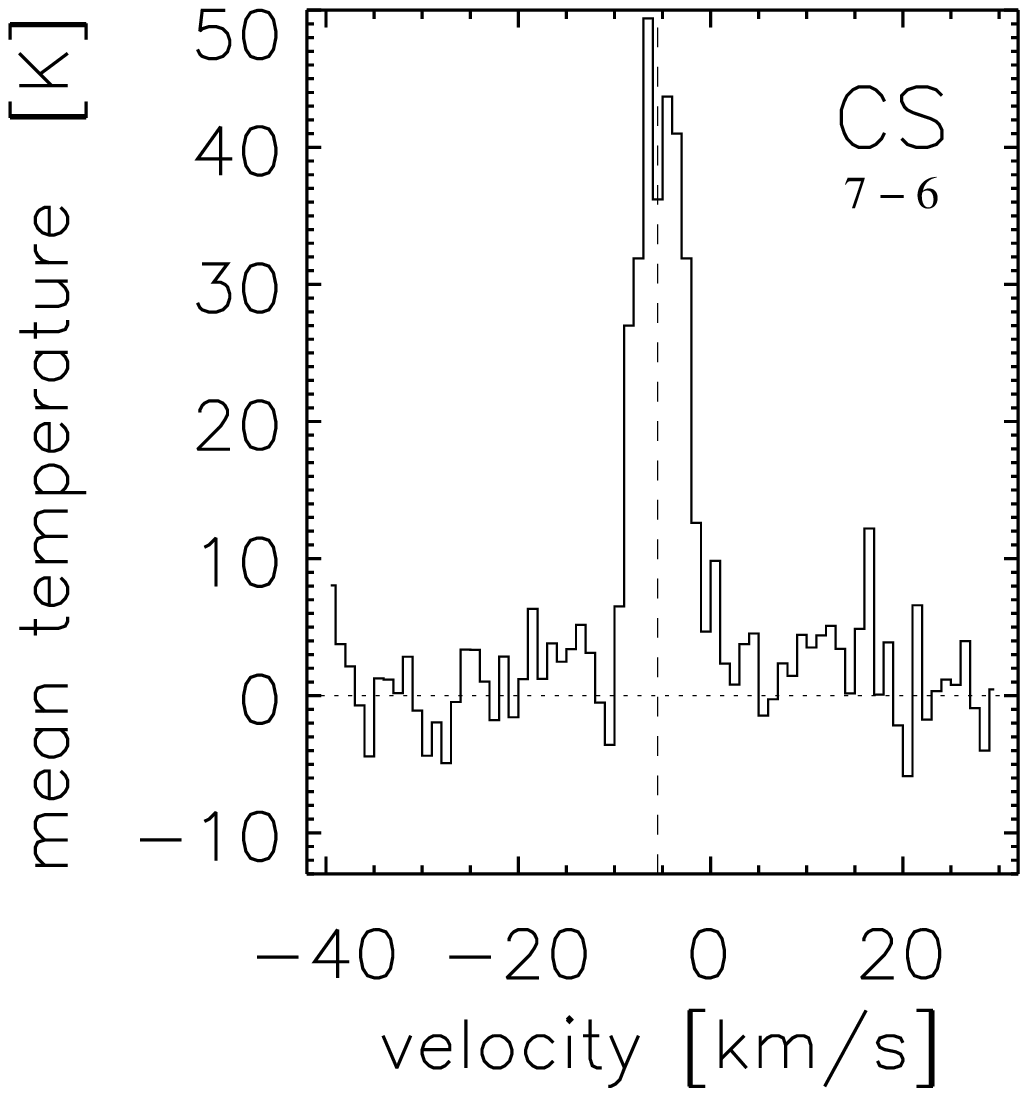}
\end{minipage}
\hspace{-0.5cm}
\centering
\begin{minipage}[t]{0.50\linewidth}
\centering
\includegraphics[width=\linewidth]{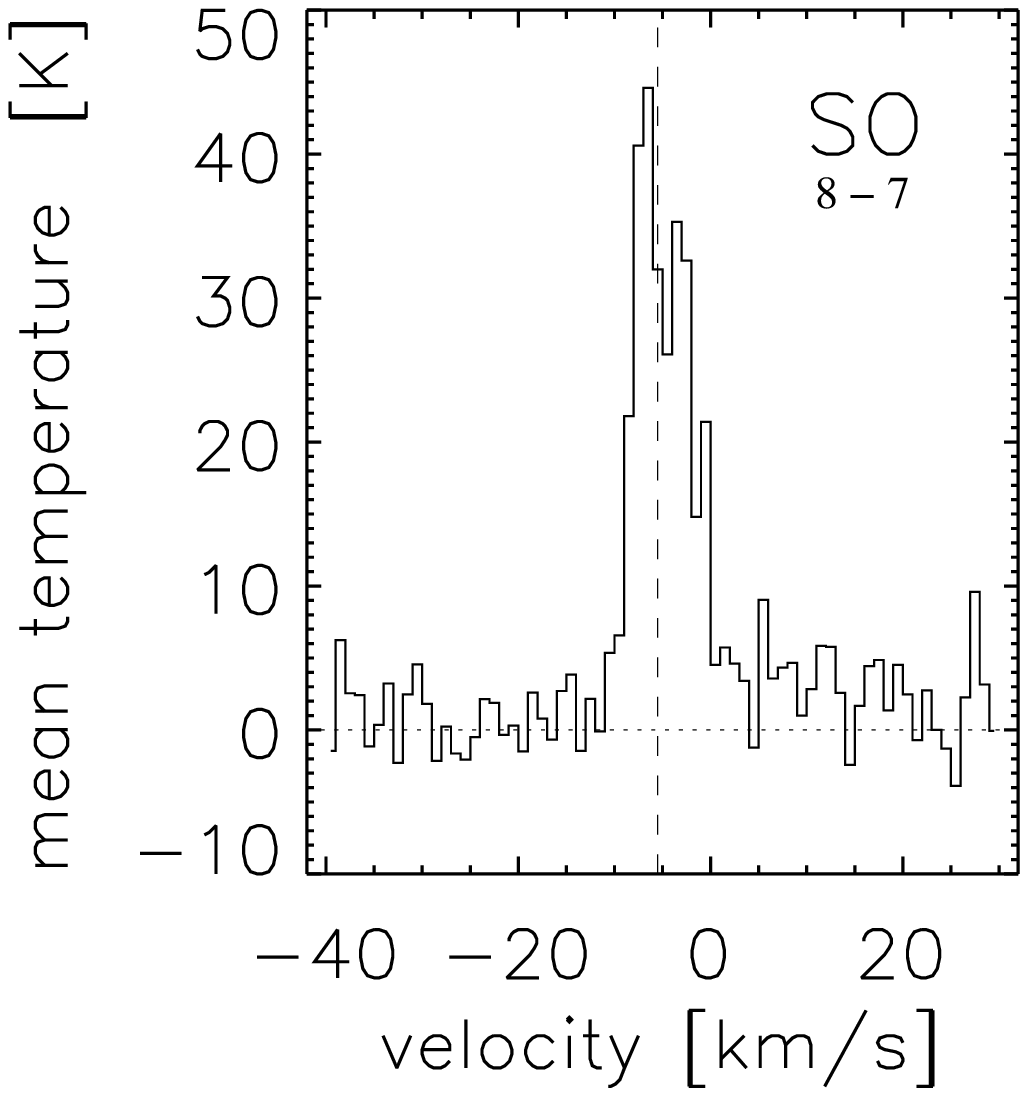}
\end{minipage}\\[-10pt]
\begin{minipage}[t]{0.50\linewidth}
\centering
\includegraphics[width=\linewidth]{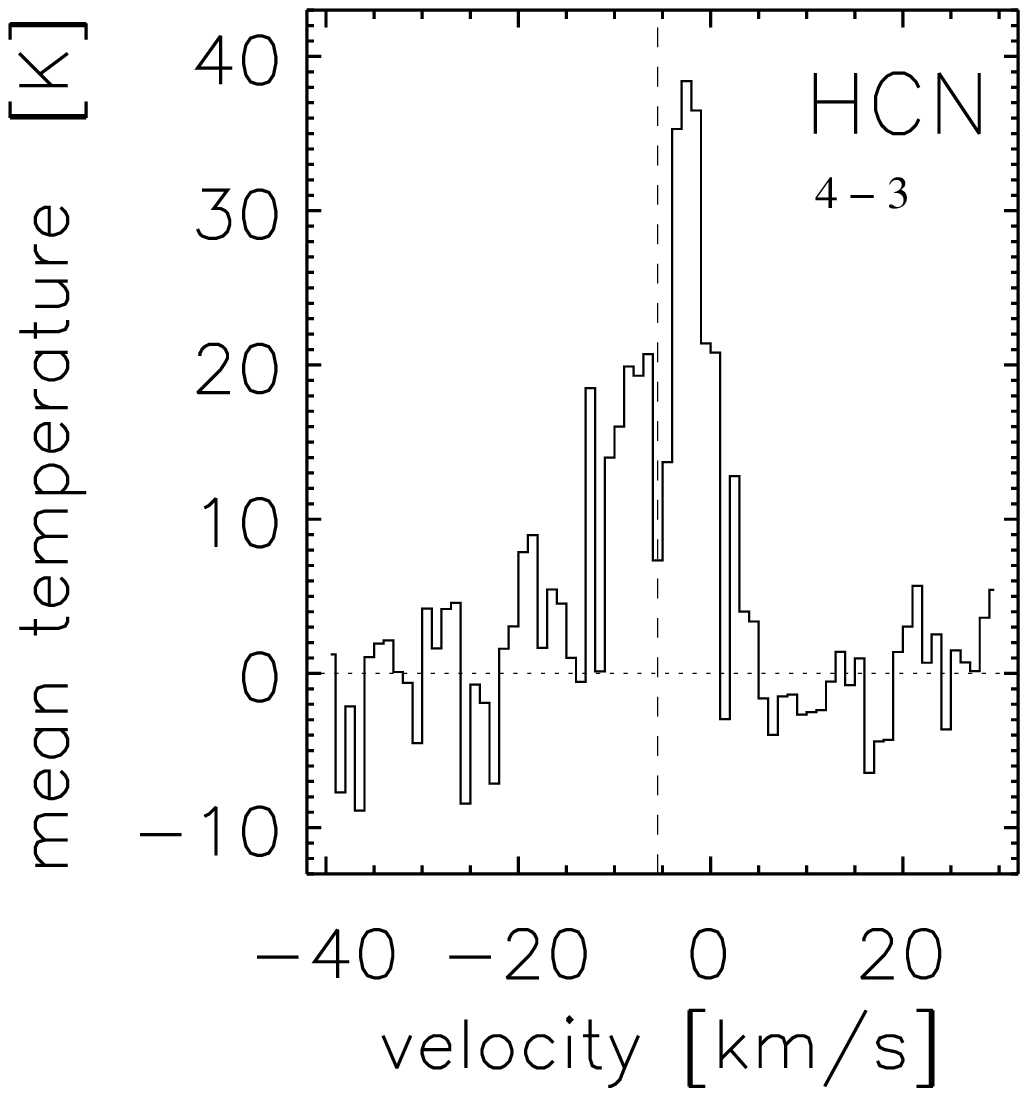}
\end{minipage}
\hspace{-0.5cm}
\centering
\begin{minipage}[t]{0.50\linewidth}
\centering
\includegraphics[width=\linewidth]{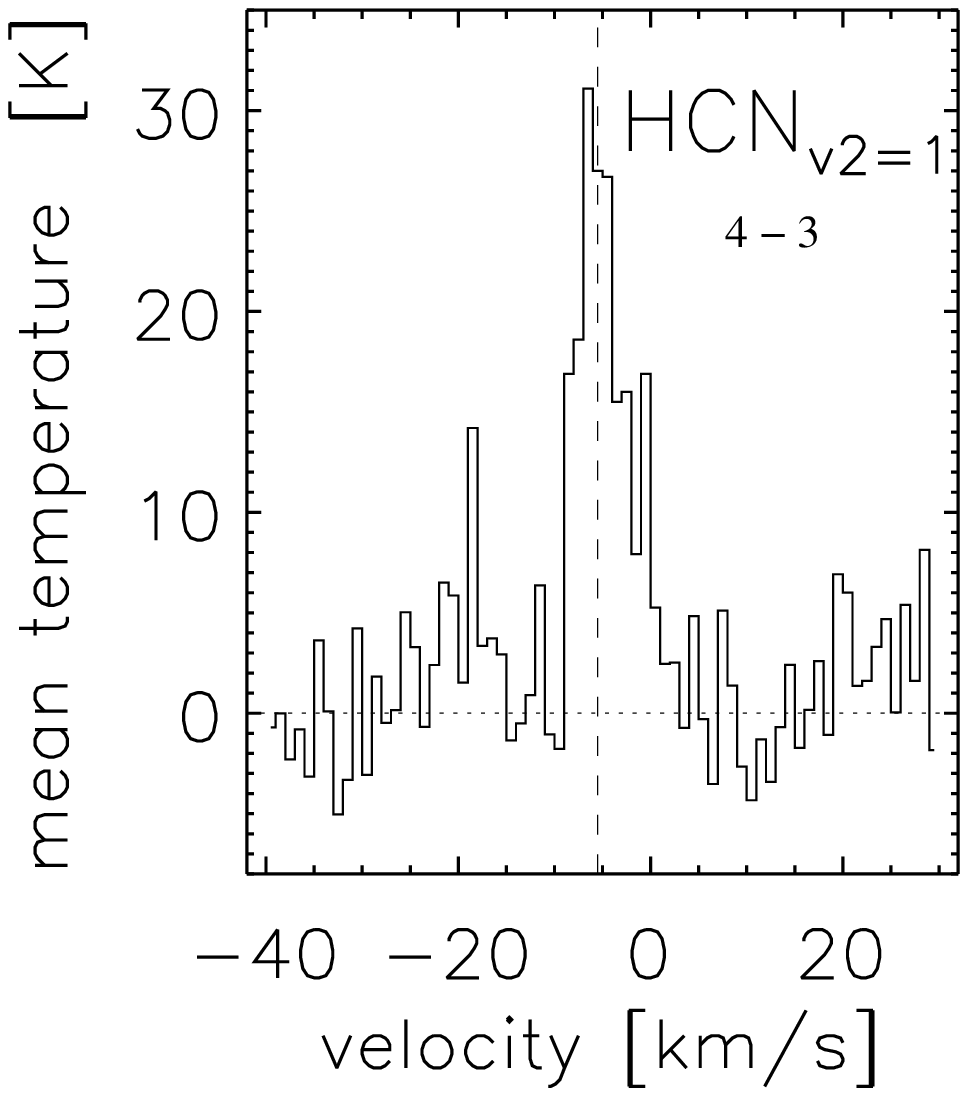}
\end{minipage}\\[-10pt]
\begin{minipage}[t]{0.50\linewidth}
\centering
\includegraphics[width=\linewidth]{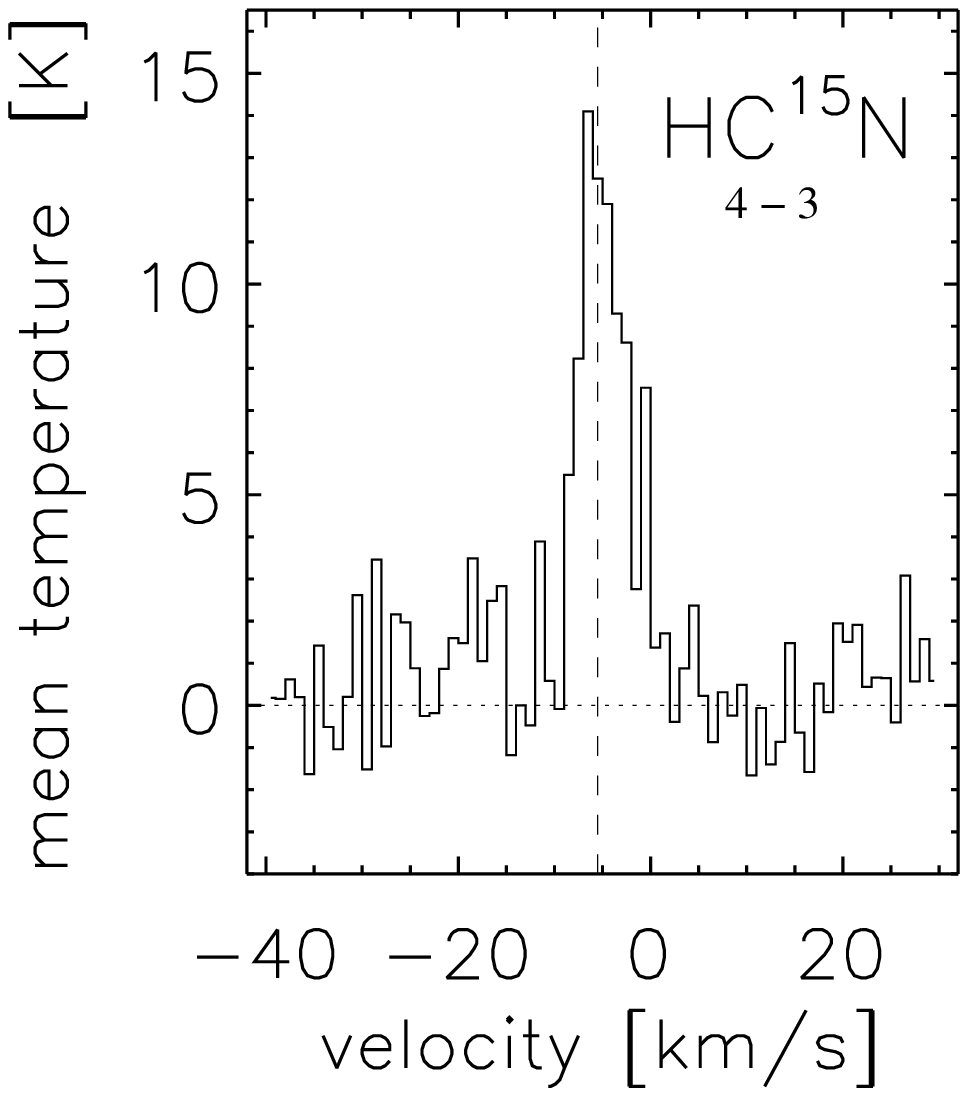}
\end{minipage}
\caption{Spectra of AFGL 2591 line emissions detected with the SMA at the
continuum peak. The brightness temperature is averaged in an area equivalent to the synthesized beam. The dashed vertical line indicates the systemic velocity of AFGL 2591.} 
\end{figure}

\begin{table*}
\begin{center}
\begin{tabular}{lcrcclrcccrcc}   
\hline \hline
Mole- &Tran-& Frequency&$E_u$&$A_{ij}$&Peak&\multicolumn{2}{c}{Line width}& 
 $V_{LSR}$&\multicolumn{2}{c}{Positional offset}&\multicolumn{2}{c}{FWHM size of axis}\\
cule &sition& & & &B.n.$^3$&\multicolumn{2}{c}{[km s$^{-1}$]}&
 &$\Delta$ RA &$\Delta$ Dec&major&minor\\
& & [GHz]&[K]&[s$^{-1}$]&[Jy/b]&SMA&JCMT& 
 [km s$^{-1}$]& [$''$]&[$''$]&[$''$]&[$''$]\\  
\hline\\
HCN&4--3&354.5055& 42.53& 2.05e-03&2.3&6.8(5)&4.60&-3.1(4)&-0.11(5)&0.01(5)&1.0(3)&0.6(2)\\
HCN& ($\nu_2$=1)4--3&354.4604&1097& 1.87e-03&3.1&6.1(1)&5.38&-5.0(3)&-0.06(4)&-0.03(3)&0.4(2)&0.3(2)\\
HC$^{15}$N&4$_1$--3$_{-1}$&344.2003&41.30& 1.88e-03&1.5&5.9(2)&4.50&-5.0(9)&-0.07(5)&0.03(4)&0.7(1)&0.7(2)\\
SO&8$_7$--7$_7$&344.3106&87.48& 5.09e-04&1.5&5.9(2)&3.70&-5.2(6)&-0.07(4)&-0.07(2)&1.4(2)&1.0(1)\\
CS&7--6&342.8830&65.83& 8.39e-04&2.1&6.1(2)&3.32&-5.8(3)&-0.02(4)&-0.06(3)&0.9(1)&0.8(2)\\
\hline
\end{tabular}
\end{center}
\caption{Observed molecules and properties in the area defined by the SMA synthesized beam centered on the maximum position of the map: peak brightness in [Jy/beam], width of Gaussian line shape fitted in velocity and shift $V_{LSR}$ with relation to the local standard of rest, positional offset of the main peak from the continuum peak, and source size. The values in parentheses indicate the error in units of the last given decimal(s).\ \  $^3$ Peak brightness of synthesized beam at position of peak continuum, using a Gaussian fit in velocity.}
\label{molecules}
\end{table*}

\section{Results of line observations}
\subsection{Spectra, maps and Gaussian fits}

The spectra of the detected lines at the continuum peak are shown in Fig. 2. There are significant differences in central velocities and line widths, but similarities in peak brightness. The line centers are consistent with the mean velocity of the envelope, $V_{LSR} = - 5.5\pm 0.2$ km s$^{-1}$ (van der Tak et al. 1999). The only exception is HCN, being red shifted by about 2 km s$^{-1}$ in the Gauss fit. A blue-shifted wing extends, however, to -13 km s$^{-1}$, possibly indicating contributions by the approaching outflow. The widths of the other lines in Fig. 2, particularly of HC$^{15}$N and SO, are relatively narrow and without compelling signatures of outflows or shocks. We note, however, that all lines are asymmetric and, except HCN, have their peaks skewed to blue velocities. It may be interpreted as the result of a high dust column depth in the innermost part of the envelope or a possible disk absorbing the blue-shifted wings. If this is the case, the red wing originates from infall. Again the main isotope HCN line is an exception and appears to be skewed to the red (Fig. 2). Note that CS, SO, and even the main isotope HCN show a dip in the spectrum exactly at the systemic velocity. It suggests that these line emissions are optically thick.

Figure 1 also displays maps of the detected lines. Images and beam size have been constructed using natural weighting. The line data were integrated from -8 to -2 km s$^{-1}$ except for HCN 4-3 where -10 to -0 km s$^{-1}$ was used. These intervals comprise roughly the  $> 3 \sigma$ intensities of the lines. The contours can be compared with the half-power size of the synthesized beam. All main peaks -- except possibly HCN($\nu_2$=1) -- appear broader than the beam, consistent with the sizes derived from two-dimensional Gaussian fits to the $(u,v)$ data, given in Tab. 1. Thus, most emission is spatially resolved by these interferometric observations. Figure 3 shows visibility amplitude plots which confirm that most of the flux is resolved out at the observed baselengths. Positional offsets and diameters of the fitted Gaussians are also given in Tab. 1. None of the differences between major and minor axes are significant, and the position angles of the source ellipses have errors that do not make the measured values statistically significant. Thus it is not possible to determine the position angle of the shape of the main peaks.

All HCN lines and CS show a secondary emission peak northwest of the main peak at about the same location. However, only in HCN($\nu_2$=1) the secondary peak exceeds 5 $\sigma$ of the background rms and is statistically significant. On the average over all the HCN and CS lines, the secondary peak is displaced from the main peak by 1180$\pm$140 AU at an angle of 21.5$^\circ$ from the RA-axis. This angle may be compared to the opening angle of the outflow as seen in near-infrared, using the speckle interferometry image of Preibisch et al. (2003). At a distance of 10 -- 20$''$, we measure a half opening angle of 30$\pm 2^\circ$ relative to the RA-axis. Thus, in projection, the secondary emission peak lies within the approaching outflow cone.

The continuum and SO line do not show enhanced emission at the position of the secondary peak. The main peak of the SO emission however is particularly extended and deviates considerably from a circular shape at the 5 $\sigma$ level. The apparent deviations of SO are to the west (approaching outflow), but possibly also to the south. These observations suggest that such deviations exist, but more observations are needed to confirm and analyze them in space and velocity. SO is the only line whose emission is marginally displaced in space from the position of the continuum emission in the southern direction. 

\begin{table*}
\begin{center}
\begin{tabular}{lrrrrrrrrrrc}   
\hline \hline
Mole-&Tran-&\multicolumn{3}{c}{Line flux}&Ratio&Ratio&\multicolumn{2}{c}{X-ray model}&\multicolumn{2}{c}{Visibility at} &Visibility at\\
cule&sition&\multicolumn{3}{c}{[Jy km s$^{-1}$]}&\underline{JCMT}&Central beam &\multicolumn{2}{c}{[Jy km s$^{-1}$]}&\multicolumn{2}{c}{$30 - 40 {\rm k}\lambda$}&$200-220{\rm k}\lambda$ \\
& &SMA&SMA&JCMT &SMA& SMA flux/Total&0.5$''$&14$''$&\multicolumn{2}{c}{[Jy km s$^{-1}$]}&[Jy km s$^{-1}$]\\  
&&central b.&total&&total&SMA flux&&&obs. &mod.& obs. \\
\hline
\\
HCN&4--3&23.3(26)&35.9&873&37.5&0.65&11.2&168&85(43)&15.6&19.2(34)\\
HCN &($\nu_2$=1) 4--3&15.2(20)&22.6&47.2&3.1&0.67&46.0&46.1&44(40)&45.4&18.6(37)\\
HC$^{15}$N&4--3&15.2(20)&23.2&59.2&3.9&0.65&17.8&18.6&45(11)&17.7&9.1(19)\\
SO&8$_7$--7$_7$&14.2(17)&35.0&151&4.6&0.41&17.6&157&80(13)&127&5.1(19)\\
CS&7--6&10.8(12)&35.9&435&40.4&0.30&11.8&143&63(14)&101&7.2(21)\\
\\
\hline
\end{tabular}
\end{center}
\caption{Observed properties of line observations. The line flux in the central SMA synthesized beam of 0.68$''\times\ 0.49''$ and total flux in the SMA image are compared with the results from van der Tak et al. (1999, 2003), who used the 14$''$ beam width of the JCMT. The corresponding values of the X-ray model (Section 5.1) are also given for 0.5$''$ and 14$''$, respectively. The visibility amplitudes in Jy km s$^{-1}$ in the intervals 30 - 40 k$\lambda$ and 200 - 220 k$\lambda$ are listed as observed and modeled.}
\label{molecules_tab}
\end{table*}

As the line emissions have significant spatial dimension in SMA images, Tab. 2 lists the fluxes contained in the synthesized SMA beam centered at the peak of the continuum emission and compares it to the total flux. The central beam refers to the brightness of the central region with a radius of approximately 300 AU integrated over the line from -15 km s$^{-1}$ to +4 km s$^{-1}$. The total line fluxes observed by the SMA are integrated in a square of 3$''\times 3''$. We found this the largest feasible area, beyond which the noise contribution dominates. The flux of the central SMA beam is a smaller fraction of the total SMA flux, the larger the source. The ratios are different for the observed molecules: low for CS and SO, higher for HCN 4-3, HCN($\nu_2$=1) and HC$^{15}$N. The small ratios of CS and SO suggest that the main CS and SO emissions have an effectively larger size, and brightness is well extending into the 600 -- 3500 AU range. This can be noticed also directly in Fig. 1, comparing CS and SO with the HCN lines. 

Table 2 furthermore juxtaposes the spectrally integrated fluxes within the central SMA beam and the JCMT beam, probing different spatial scales. SMA fluxes are between 3 and 40 times smaller than the JCMT values measured by a beam that is 530 times larger in area. If the emission came from a peak unresolved in both beams, the fluxes would be the same. The brightness ratio of HCN($\nu_2$=1) is lowest, suggesting that the source of this line is not much larger than the SMA beam, as noticed before (Fig. 1 and Tab. 1). The JCMT fluxes for the main HCN isotope and CS are more than 30 times larger, indicating that their main contribution originates from the envelope at several thousand AU, the central peak being probably limited by optical thickness. 

The widths (FWHM) of the velocity distributions are also given in Tab. 1. In the JCMT field of view (14$''$ diameter), the line widths are relatively uniform in the range of 4.3$\pm$1.0 km s$^{-1}$. The values measured by SMA, referring to a region more than an order of magnitude closer to the YSO are around 6 km s$^{-1}$. HCN($\nu_2=1$) is noteworthy as it changes the least. As HCN($\nu_2=1$) is very compact in the SMA observations, this can again be interpreted in terms of the emission being dominated by a centrally peaked component. It adds to the claim that for this line the JCMT is picking up mostly the same flux seen by the SMA (Tab. 2). We note for comparison that the free-fall velocity from infinity to 7000 AU, relevant for the JCMT beam, amounts to 2.6 km s$^{-1}$. At 500 AU corresponding to the SMA synthesized beam, it is 7.5 km s$^{-1}$, neglecting the mass of the inner envelope and disk. 

\begin{figure*}
\centering
\begin{minipage}[t]{0.38\linewidth}
\centering
\includegraphics[width=\linewidth]{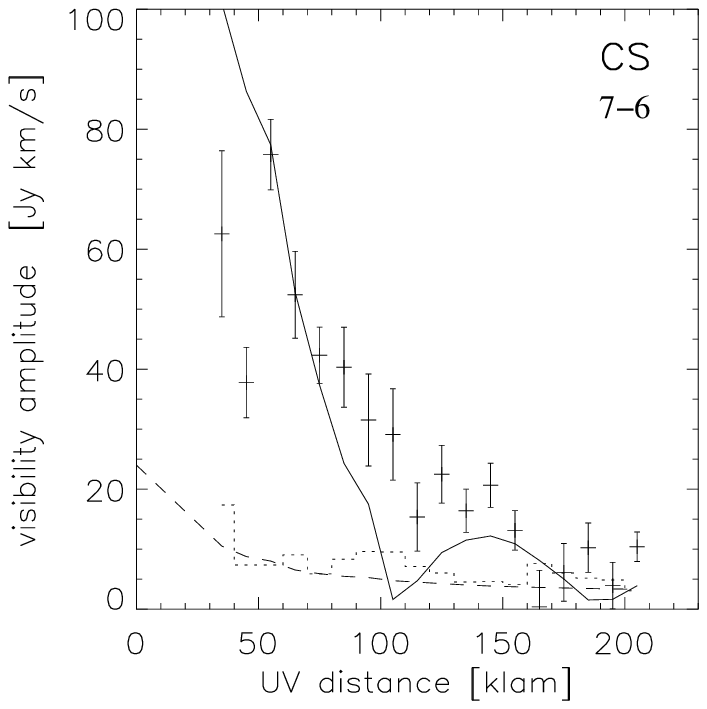}
\end{minipage}
\hspace{0.4cm}%
\centering
\begin{minipage}[t]{0.38\linewidth}
\centering
\includegraphics[width=\linewidth]{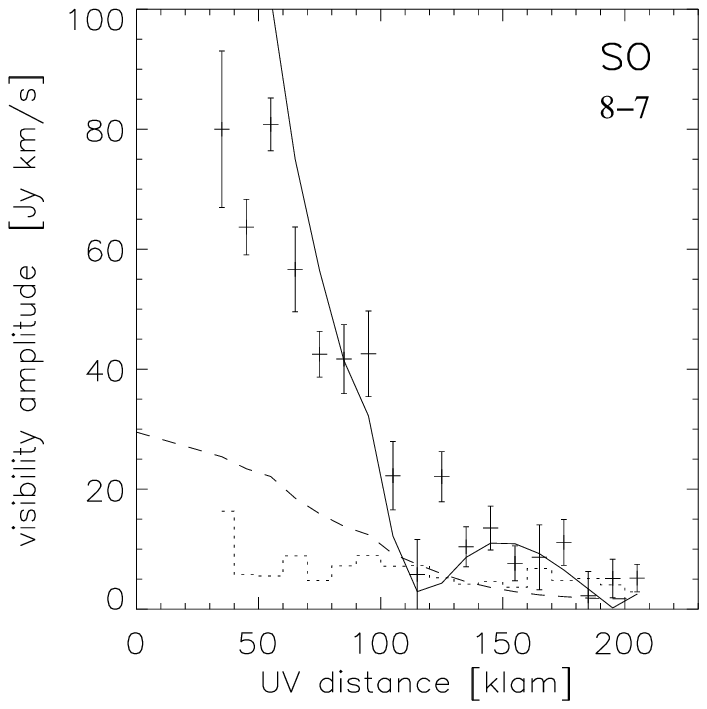}
\end{minipage}\\[20pt]
\begin{minipage}[t]{0.38\linewidth}
\centering
\includegraphics[width=\linewidth]{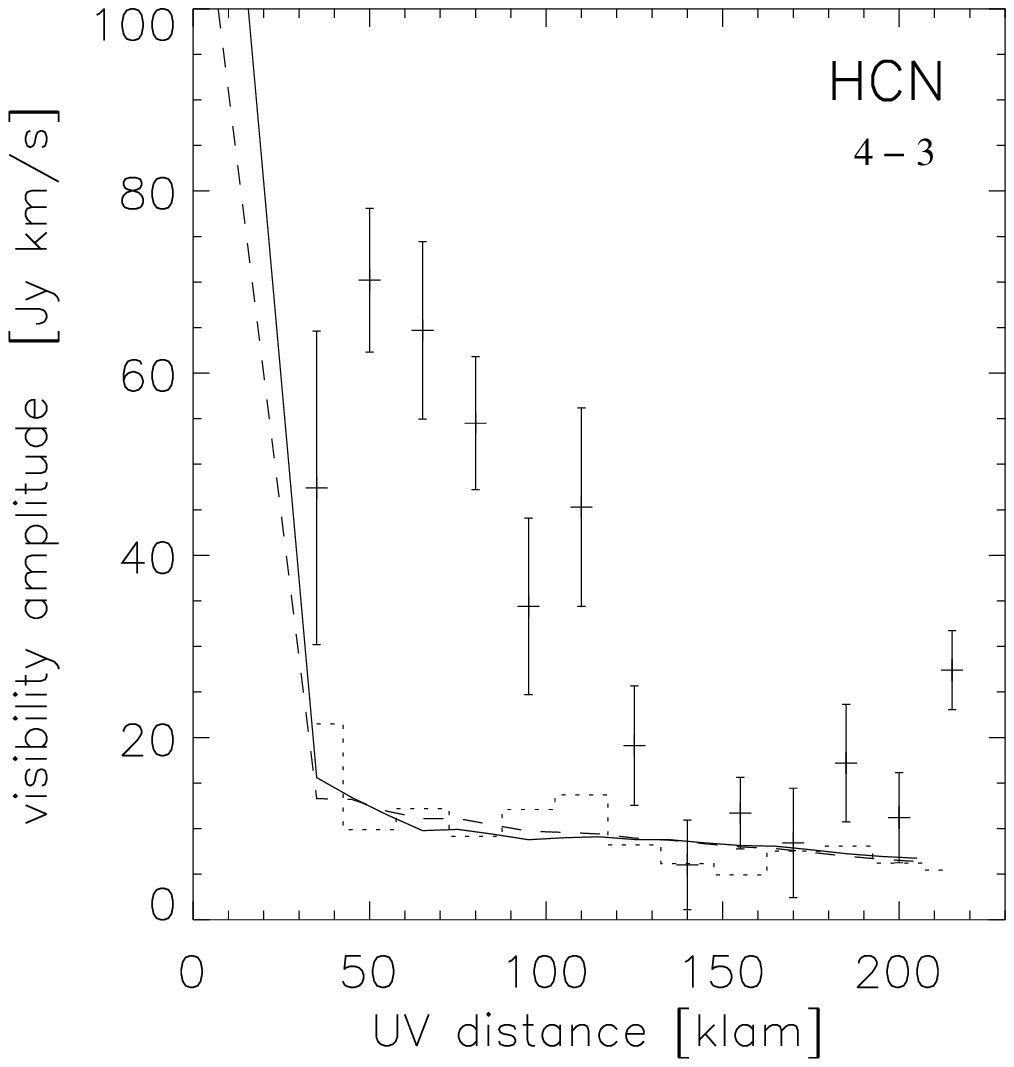}
\end{minipage}
\hspace{0.4cm}%
\centering
\begin{minipage}[t]{0.38\linewidth}
\centering
\includegraphics[width=\linewidth]{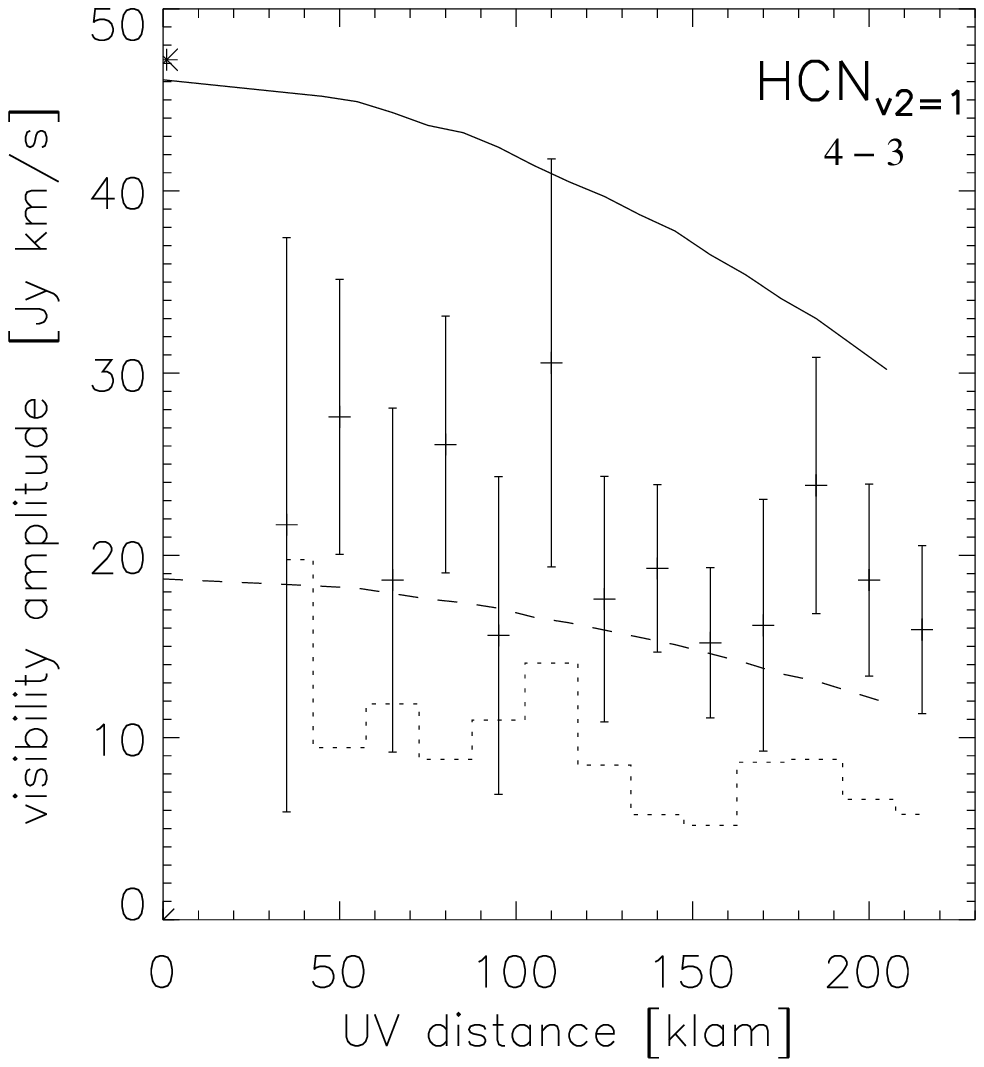}
\end{minipage}\\[20pt]
\begin{minipage}[t]{0.38\linewidth}
\centering
\includegraphics[width=\linewidth]{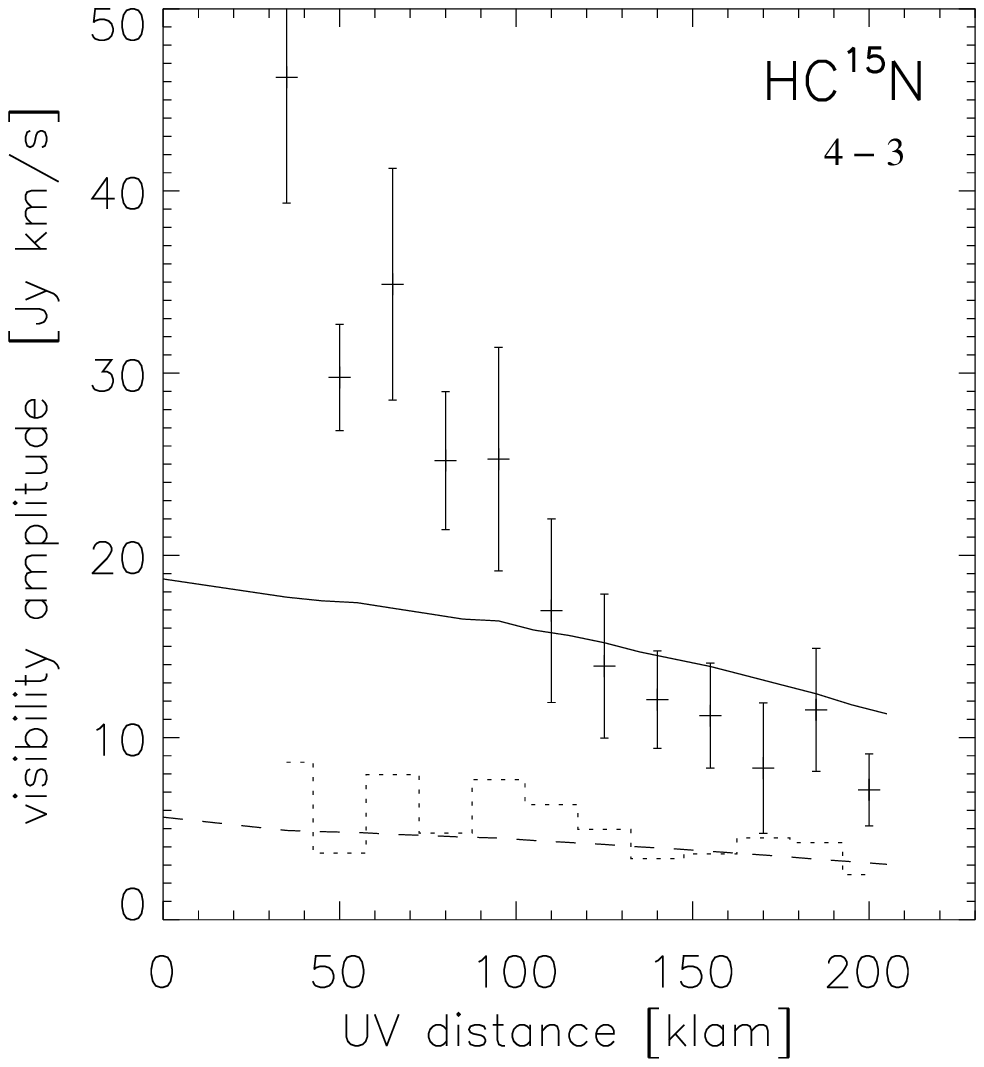}
\end{minipage}
\hspace{0.4cm}%
\centering
\begin{minipage}[t]{0.38\linewidth}
\centering
\includegraphics[width=\linewidth]{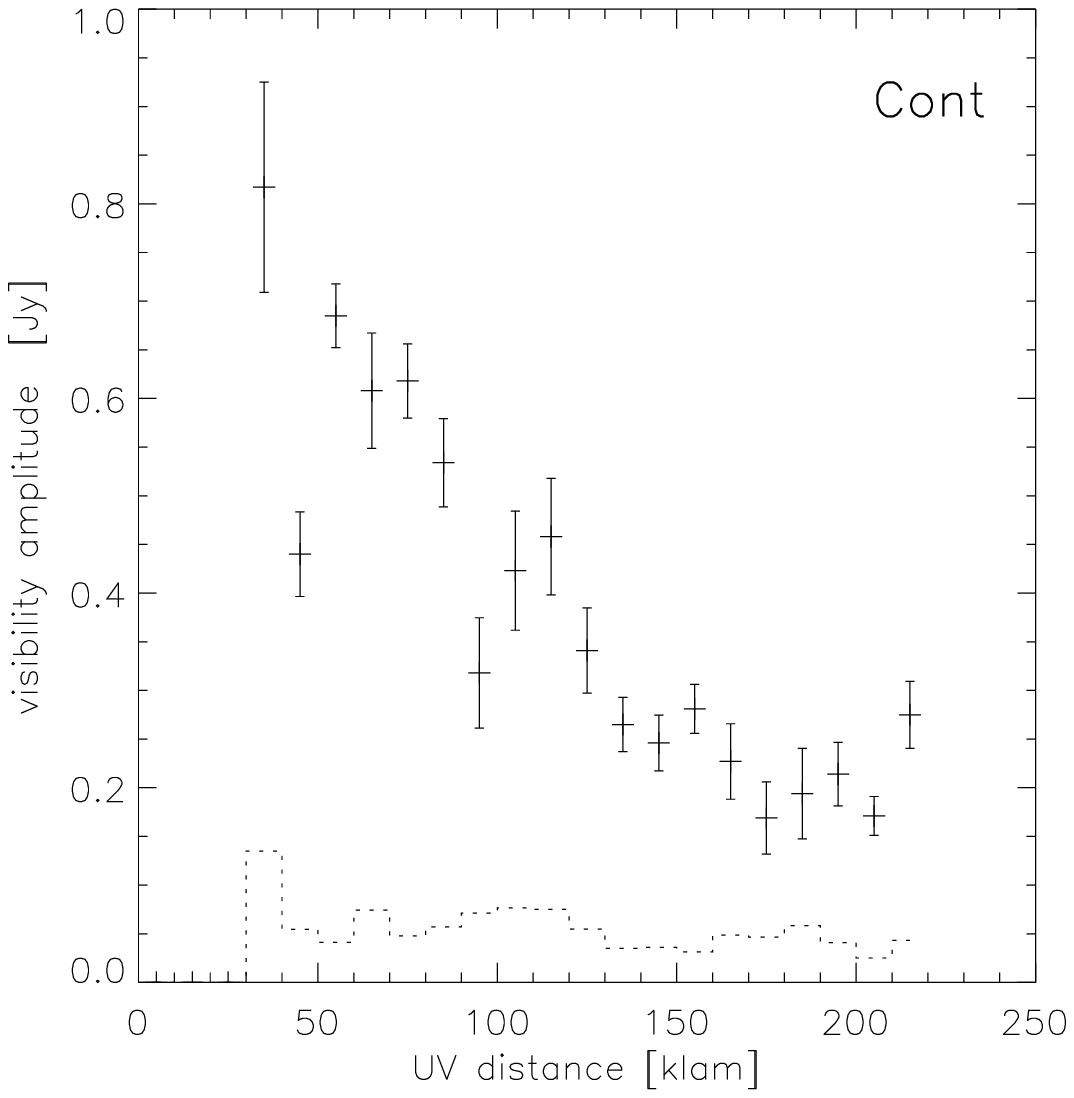}
\end{minipage}
\caption{Visibility amplitudes of AFGL 2591 in the observed line emissions vs. projected baseline distance in units of $10^3 \lambda$. The observed values are shown with {\it error bars}. They were integrated from -9.5 to -1.5 km s$^{-1}$ except for HCN 4-3 where -10.5 to -0.5 km s$^{-1}$ was used. The {\it dotted histogram} indicates the expected noise level assuming no signal. The {\it solid curve} represents the model (introduced in Section 5), assuming protostellar far-UV and X-ray emission. The model without X-rays is shown {\it dashed}. The single-dish values measured at JCMT are beyond the scale except for HCN($\nu_2$=1), where it is indicated by an asterisk at zero $(u,v)$ distance. The visibility amplitudes of the continuum emission (integrated in space, Fig. 3f) are given in flux density, averaged over the observed frequency range.} 
\end{figure*}

\subsection{Brightness distributions (visibilities)}
The visibility amplitudes vs. $(u,v)$ distance, representing the brightness distributions in space are presented in Fig. 3. The $(u,v)$ distance, $x_{k\lambda}$, corresponds to the baseline length $b$ in units of the observed wavelength $\lambda$ projected to the plane of the sky at the peak position. It is related to the angular structure by the Fourier transformation of the image. The angular fringe separation $\Delta\theta$ is given by 
\begin{equation}
\Delta \theta \ \ =\ \ {\lambda\over {b \sin \theta}}\ =\ \ {206.26\over x_{k\lambda}}\ \ \ \ {\rm [arcsec].}
\label{Delta}
\end{equation}  
When discussing spatial structures, it is more convenient to use angular radii $r$, where $2 r=\Delta\theta$, and convert to AU (1$''$ = 1000 AU for our assumed distance to AFGL 2591). Each baseline yields an average value in one dimension. In spherical symmetry, the average source diameter is $\bar d = 0.5 \pi r$. Thus the relevant radius in the model is related to the one-dimensional resolution by
\begin{equation}
r\ \ =\ \ {2 \Delta\theta\over\pi}\ \  =\ \ {131.31\over x_{k\lambda}}\ \ \ {\rm [1000\ AU].}
\label{radius}
\end{equation}  

The visibilities at the longest and shortest baselines yield information on the smallest observable spatial dimension. At the longest baselines ($>$ 200 k$\lambda$, see Fig. 3), all observed lines, except for SO and possibly the main isotope HCN, show visibility amplitudes  
exceeding the noise. This indicates that the map contains a peak, most likely in the center, that is partially unresolved ($<$ 600 AU in radius). Table 2 gives the average value in the range 200 - 220 k$\lambda$, equivalent to the unresolved flux. As HCN($\nu_2$=1) and HC$^{15}$N have an unresolved component, its relatively low value in the main HCN isotope is likely an optical depth effect. 

\begin{figure*}
\centering
\resizebox{15cm}{10cm}{\includegraphics{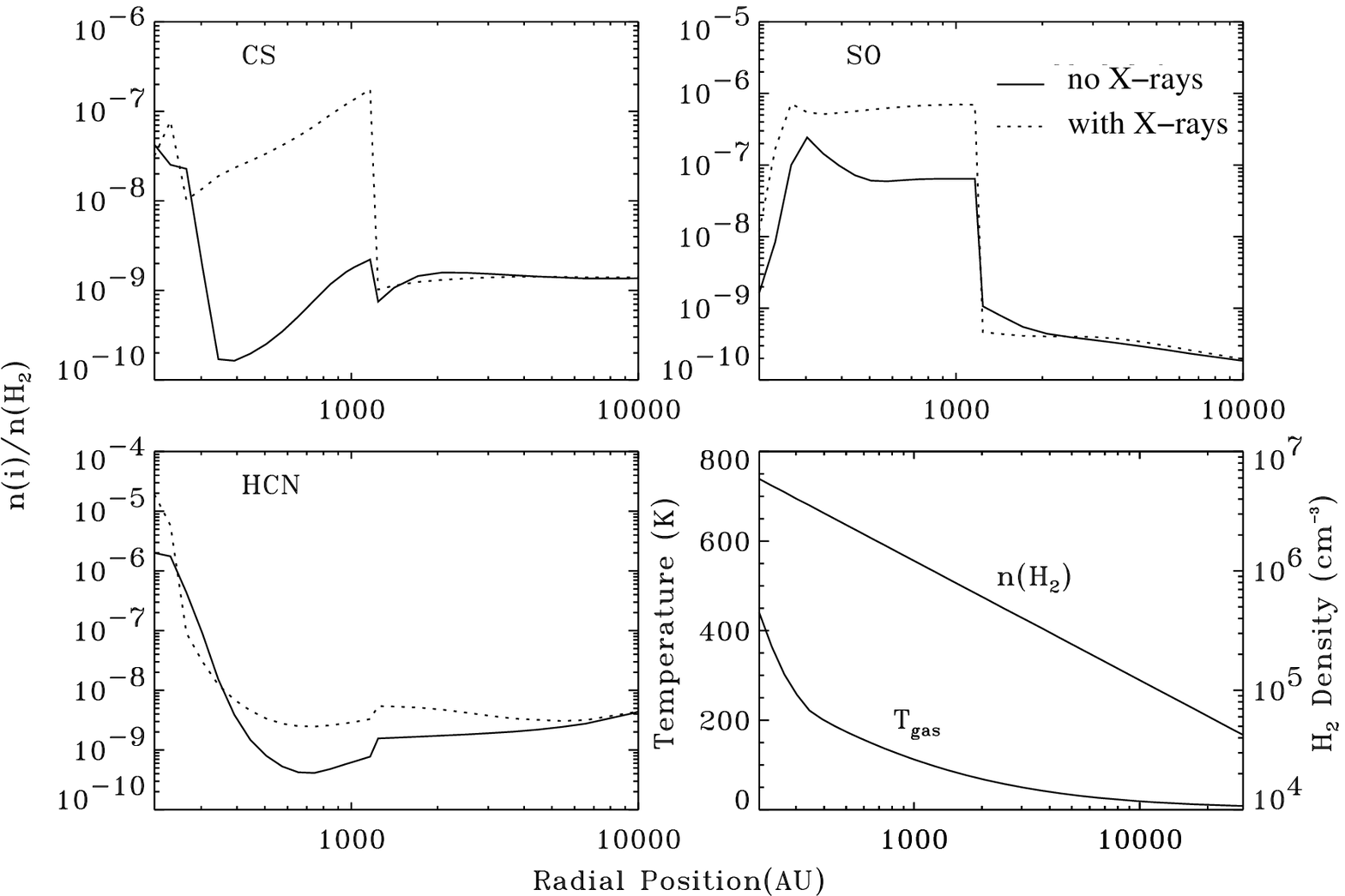}}
\caption{{\sl a - c:} Radial models of chemical abundance as derived by St\"auber et al. (2005). The {\it solid curve} shows the model of the envelope without far-UV and X-ray irradiation by the central object (no-X model), and the {\it dotted curve} shows the X-ray model including an X-ray luminosity of $8\cdot 10^{31}$ erg s$^{-1}$ and protostellar far-UV irradiation of $G_0=10$ at 200 AU fitting best to the single-dish data on 24 molecules. {\sl Fig. 4d:} Density and temperature model of AFGL 2591 used for chemical modeling (see text).} 
\label{models}
\end{figure*}

\section{Models}
The visibility amplitudes of lines contain information on the radial variations of molecular abundances. In this section, the tools and models are presented that will be used in the following section to compare theory and observations. First, the results of chemical modeling are shown, assuming a spherically symmetric density and temperature envelope model derived from observations of AFGL 2591. The chemical models include ice evaporation and assume different levels of far-UV and X-ray irradiation. The models predict visibility amplitudes that will be compared to observations in the following section. 

Models of the fractional abundances of CS, SO and HCN are shown in Fig. 4a-c. Molecular abundances are calculated in a time-variable chemical network including 395 molecular species and 3865 reactions (Doty et al. 2002). The input density model (Fig. 4d) is based on a $1/r$ power-law distribution derived by van der Tak et al. (1999), and the input temperature model is from Doty et al. (2002). Abundances are determined in time and radial distance. A first model, which we call the 'no-X' model in the following, fits best 24 column densities derived from observations under the assumption of no internal X-ray or far-UV radiation. The deviations reach a minimum for a chemical age of 5$\cdot 10^4$yr (Doty et al. 2002). St\"auber et al. (2005) expanded the model to allow for irradiation by X-rays and far-UV and searched for the best fit for the same molecular observations, allowing for irradiation by a central source. The fitted parameters are X-ray luminosity, the temperature of the X-ray emitting plasma, far-UV irradiation at 200 AU, and age. The result, which we call 'X-ray model' although it also includes far-UV irradiation, requires $L_x = 8\cdot 10^{31}$ erg s$^{-1}$, 10$^8$ K, $G_0=10$, and 5$\cdot 10^4$yr, respectively. $G_0$ is the scaling factor of the average interstellar far-UV radiation flux ($G_0$=1 corresponds to 1.6$\cdot 10^{-3}$ erg cm$^{-2}$s$^{-1}$ at 6$<h\nu<$13.6 eV). Comparing modeled and molecular column densities observed by single dishes, the X-ray model has a significantly smaller $\chi^2$ deviation than the no-X model (St\"auber et al. 2005).

The dominant form of sulfur in the grain mantles depends on the gas density of the pre-stellar phase. Low density favors hydrogenation and H$_2$S (Charnley 1997), and high density leads to oxygenation (OCS, van der Tak et al. 2000). Laboratory results by Collings et al. (2004) suggest that the sulfur-bearing molecules are small enough to diffuse through the structure of porous amorphous water ice and desorb at intermediate temperatures between 100 K (H$_2$O) and 20 K (CO) under interstellar conditions. In the following we refer to the location of H$_2$S and OCS evaporation as the 'sulfur evaporation radius'. Within this radius, sulfur resides in gas-phase molecules, rapidly undergoing reactions that create CS, SO, SO$_2$ and other sulfur-bearing molecules. Outside of the sulfur evaporation radius, sulfur is assumed to be mostly frozen out on grains, and the total gas-phase abundance of sulfur is only $3\cdot 10^{-9}$, maintained by cosmic ray sputtering. The abundances of these molecules will thus show a jump at the sulfur evaporation radius from a high value inside to a very low level farther out. Chemical modeling predicts that the abundances of CS and SO in the gas phase are enhanced by X-rays after sulfur evaporation. According to the model of Doty et al. (2002), the $T = 100$ K surface has a radius of 1300 AU. Doty et al. (2006), on the other hand, predict the water evaporation edge at about 670 AU, assuming their preferred age of $8\cdot 10^4$ yr since YSO formation and using a dynamical model. The magnitude of the jump depends strongly on X-ray irradiation. In the X-ray model the jump amounts to two orders of magnitude for CS and to three orders of magnitude for SO.
 
HCN is little affected by sulfur and water evaporation. Its abundance is greatly enhanced, however, in the central region (Sect. 1 and Fig. 4c) due to the presence of atomic N and C, and the lack of atomic O. This situation arises through the following processes: At temperatures above 230 K, the reaction with H$_2$ drives most atomic oxygen into OH and finally H$_2$O (Charnley 1997; Rodgers \& Charnley 2001). On the other hand, atomic nitrogen is available in the central region from the dissociation of N$_2$ by far-UV photons. The association of N by the reaction NO + N $\rightarrow$ N$_2$ + O is suppressed as all oxygen resides in H$_2$O. Thus atomic N and C are available and form CN. As CN + H$_2$ $\rightarrow$ HCN + H has an activation energy of 820 K, CN is hydrogenated to HCN only at high temperatures. In regions not irradiated by X-rays and far-UV photons, atomic carbon forms through photodissociation of CO induced by cosmic rays. If protostellar far-UV or X-ray irradiation enhance the C abundance, HCN is formed more numerously. Therefore, the HCN abundance is greatly enhanced by irradiation at high temperatures as modeled in the central region within a radius of 250 AU. In outflows or outflow walls direct far-UV irradiation and heating may enhance HCN. This is a possible cause of the secondary peaks.

To find the visibility amplitudes of the models as observed by the SMA, the molecular density is first determined by folding the fractional abundance with the density model of H$_2$ (Fig. 4d). The 1D Monte Carlo code of Hogerheijde \& van der Tak (2000) and molecular data from the Leiden atomic and molecular data base (Sch\"oier et al. 2005) are then used to determine the molecular level populations. Then, radiative transfer is calculated. The result is integrated along each line of sight into a two dimensional map (Fig. 5). Optical depth effects are included, but not non-radial inhomogeneity or velocity flows. The map is finally transformed into visibility amplitudes applying the $(u,v)$ coverage of the actual SMA observations with the help of MIRIAD. 

The modeling of ro-vibrational HCN($\nu_2$=1) emission requires some assumptions. The level populations were approximated from the rotational 4-3 transition and weighted by the density,
\begin{eqnarray}&n({\rm {HCN}}(\nu_2&=1)) = {n({\rm {HCN}})g_u({\rm {HCN}}(\nu_2=1))\over g_u({\rm {HCN}})} \nonumber\\
& & \times\ \exp\left({-[E_u({\rm {HCN}}(\nu_2=1))-E_u({\rm {HCN}})]\over kT}\right) \ \ \ ,
\end{eqnarray}
assuming that the levels are thermalized and in LTE. Thus the levels $\nu_2=0$ and 1 have similar abundances at high temperatures, and the ratio is close to zero at $T<20$ K. 

To discuss the effects of abundance variations in radius on visibility amplitudes, we recall four basic properties of Fourier transformation. ({\it i}) Adding a small quasi-constant component to the abundance enhances just the zero value and leaves unchanged the visibilities at non-zero baselines. ({\it ii}) The visibility amplitudes of a partially resolved peak depend linearly on the flux of this source. ({\it iii}) Expanding a resolved Gaussian source in size reduces the non-zero amplitudes. A jump in the density acts like an additional source having the size of the jump radius and enhances the visibilities below the corresponding baseline (Eq. 2). ({\it iv}) A boxcar density model transforms into a $sinc$-function in visibility amplitude. A boxcar molecular abundance distribution out to a radius $r$ thus causes nulls in visibility amplitude at $x_{k\lambda}$ given by Eq. (2). These properties of Fourier transformation persist basically in the case of a power-law given by the density model. An abundance jump at 1300 AU predicted by the model for CS and SO thus yields a zero in visibility amplitude at 113 k$\lambda$. The value can differ slightly if the density model does not have an exact boxcar shape. Visibilities observed for AFGL 2591 in the range of 38 to 215 k$\lambda$ are sensitive indicators of spatial structures with radii from a few hundred to a few thousand AU. 

\begin{figure}
\centering
\resizebox{\hsize}{!}{\includegraphics{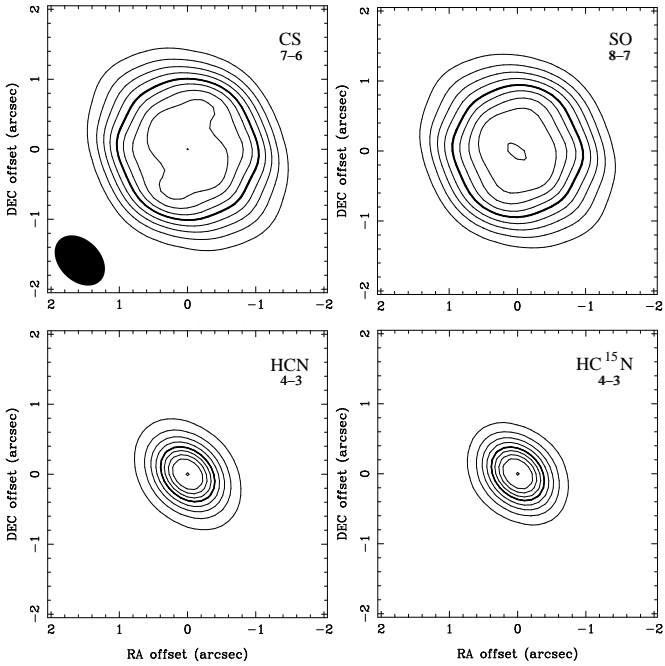}}
\caption{Line maps of chemical model including X-rays and far-UV irradiation as if observed by the SMA. Contours are at 10\% steps of the peak. The 50\% curve is {\it bold}. The SMA beam is shown for comparison in Fig. 5a.} 
\label{model maps}
\end{figure}

\section{Comparison of models with data}
There are four ways to compare the chemical model of St\"auber et al. (2005) with the data: 
\begin{itemize}
\item The total flux of the model in a 14$''\times 14''$ area can be readily compared with the single-dish observations by the JCMT (both given in Tab. 2). The total JCMT flux corresponds to the value at zero k$\lambda$ in Fig. 3 and includes in particular the radii 3500 -- 7000 AU, missing in the SMA total flux. As the chemical model was fitted to 24 molecular column depths derived from single-dish data including the lines selected here, it is not surprising to find here general agreement. The largest deviation is the main isotope HCN, observed by the JCMT more than five times brighter than predicted.  
\item The region with radii 600 -- 3500 AU is best compared in visibility amplitudes (Fig. 3). It is of special interest as it contains the region influenced by protostellar X-ray emission as well as the sulfur evaporation region.
\item The center region ($<$ 600 AU) is related to the largest $(u,v)$ distances (200 -- 220  k$\lambda$, Tab. 2). Alternatively, the 0.5$''$ central field of the model can be tested by the central SMA beam measurement (marked 0.5$''$ in Tab. 2). The flux observed in the SMA beam at the peak of the continuum agrees with the unresolved flux within the given accuracy except for the flat SO emission (see next paragraph). 
\item Finally, the predicted size of the model depicted in Fig. 5 can be contrasted to the observed map (Fig. 1) and the Gauss fit (Tab. 1).
\end{itemize}

A first, important result: The molecules reacting to sulfur evaporation, CS and SO, have larger emitting regions than the HCN lines at the 50\% isophote in Fig. 1 as expected from the X-ray model (Fig. 5). CS has a smaller diameter measured by the Gauss fit than SO (Tab. 1). In effect, the CS emission contains less flux in the central peak and has a higher peak. Thus its shape is spikier than the SO emission. CS appears clearly more extended than HCN. In the following we distinguish between the size at the FWHM level and that of a broad pedestal, and discuss the lines individually. 

\subsection{CS line emission}
In general, the modeled CS visibilities agree with the observed ones. There are three deviations: {\it (i)} The 0 k$\lambda$ value is a factor of 3 below the observed one (JCMT flux, Tab. 2). This indicates that the model underestimates the flux at radii 3500 -- 7000 AU. {\it (ii)} The CS model produces marked minima at 105 and 190 k$\lambda$ where the amplitude reaches nearly zero (Fig. 3a). It is the result of the extremely sharp abundance jump by two orders of magnitude at the sulfur evaporation edge (X-ray model, Fig.4a), causing a null in Fourier space at 102 k$\lambda$ and its approximate integer multiples. {\it (iii)} Figure 5 predicts a FWHM size of about 2.1$''$, remarkably larger than 1.04$''$, as measured from the main emission peak (Tab. 1) convolved with the beam. 

\subsection{SO line emission}
The model for SO 8$_8$-7$_7$ emission also agrees relatively well with the data. In particular, we note that the 0 k$\lambda$ value agrees with the observed one (JCMT flux, Tab. 2). Most importantly SO being our best X-ray tracer, the X-ray model reproduces the observed data much better than the no-X model. The SO abundance decreases below 300 AU (Fig. 4b), reducing the optical thickness in the central region and diminishing the Fourier power of the unresolved central region. Thus the SO visibilities at $>$200 k$\lambda$ are predicted to be relatively low and less peaked than CS. Indeed, the observations show no signature of an unresolved central peak contrary to CS. 

There are two major deviations: {\it (i)} Like for CS, the model implies a jump in SO abundance at 1300 AU by more than a factor of 1000 (Fig. 4b). The deep minima in Fourier space are not observed. Nevertheless, a dip in the observed visibilities at 115 k$\lambda$ is suggestively close to the predicted 117 k$\lambda$ (Fig. 3b). If real, it may be interpreted by a jump in abundance. The dip may be visible only in SO where the jump is predicted to be larger than in CS. {\it (ii)} On the other hand, the predicted FWHM size is 1.9$''$ (Fig. 5), a factor of about 1.6 larger than observed (Tab. 1, Fig. 1). If the size were determined by a jump, the dip would be expected at 1.6 times smaller $(u,v)$ distance, again contrary to observations (Fig. 3). 

\subsection{HCN line emissions}
The main isotope HCN 4-3 model underestimates the total flux at the 7000 AU radius scale by a factor of 5. Thus HCN has an additional intense pedestal emission collected by the JCMT single dish, but not predicted by the model. On the other hand, the model fits roughly in the inner part ($<$ 1100 AU, corresponding to $>$ 120 k$\lambda$). The predicted FWHM of 0.8$''$ (Fig. 5) agrees with observations. 

Contrary to the rotational HCN line, the modeled ro-vibrational HCN($\nu_2$=1) total flux agrees well with the JCMT observations. However, the visibility amplitudes over the range 60 to beyond 210 k$\lambda$ (radii 600 -- 2200 AU) are overestimated, suggesting a main peak size larger than modeled by about a factor of 3. This is corroborated by the flux in the central SMA beam, being smaller than predicted (Tab. 2). 

The HC$^{15}$N model follows the main isotope HCN in many properties. It fits relatively well with the data in the inner part, predicting even the unresolved peak flux to within better than a factor of two. The model fits well at large visibilities (Fig. 3, $\approx$ 600 AU), and in the unresolved peak. There is also good agreement with the observed FWHM size. However, the same shortcoming persists: The total flux is a factor of 3 lower than observed by the JCMT single-dish. It requires an additional extended component at radii $>$ 1100 AU ($<$ 120 k $\lambda$). 

In conclusion, HCN and HC$^{15}$N agree qualitatively with the predicted central abundance peak, but require an additional broad pedestal emission beyond about 1100 AU.  This broad component is not clearly perceptible in HCN($\nu_2$=1) visibilities, although the secondary emission peak is detected also in this line (Fig. 1). 

\section{Discussion}
\subsection{X-ray and far-UV irradiation}
The question of X-ray emission by the deeply embedded HMPO AFGL 2591 can be addressed by these SMA observations in more depth than previous studies based only on single-dish observations. Spatial resolution allows testing independently if and how X-rays affect molecular abundances with distance from the protostar, depending on density, temperature and irradiation. The X-ray model of St\"auber et al. (2005) includes also far-UV irradiation, which has little effect beyond 300 AU in a spherically symmetric model. If present, X-ray and far-UV irradiation have similar chemical effects, differing in the primary ionization and in penetration depth. In the inner part ($<$ 1000 AU) of a spherically symmetric model, SO mainly traces X-rays, and HCN mainly traces far-UV irradiation (Sect. 1). CS can be enhanced by both irradiations (St\"auber et al. 2004, 2005).

These SMA observations support the hypothesis of X-ray irradiation in two ways: Firstly, Fig. 3b demonstrates that the model including X-rays fits the SO visibility amplitude better than the no-X model. The difference is remarkable at large radii ($\gapprox$ 1300 AU, corresponding to $\lapprox 100$ k$\lambda$) where X-rays completely dominate far-UV irradiation. Secondly, the CS emission is predicted to have a large main source in the presence of X-rays. Figure 4a shows that this is the result of X-ray irradiation within the sulfur and water evaporation region and does not occur without X-rays. At radii larger than about 1000 AU, the CS source size becomes therefore a further X-ray tracer. The measurements are consistent with this prediction and thus corroborate X-rays.

The HCN abundance is sensitive to far-UV irradiation, more than to X-rays (Section 1). The HCN lines deviate from the modeled visibility amplitudes mostly in the 500 -- 3500 AU radii range (Fig. 3) at distances comparable to the secondary emission peak possibly located in the outflow. In addition, HCN and HC$^{15}$N, but not HCN($\nu_2$=1), contain a broad component beyond 3500 AU accounting for nearly 90\% of the main isotope HCN emission. The HCN($\nu_2$=1) line is preferentially emitted in the very inner part. This appears to be the result of its excitation mechanism. It is supported by the high brightness temperature of the HCN($\nu_2$=1) line (Tab. 1), consistent with observed HCN absorption lines in the infrared indicating temperatures up to 1300 K (Boonman et al. 2001) and its small source size. 

Far-UV radiation propagates only in a low density gas, but scatters more easily than X-rays. The deviations of HCN and the CS from the predictions of a 1D envelope model may be explained as additional emission in outflows or outflow walls. HCN emission in high-mass outflows has recently been reported by Zhang et al. (2007). In addition to far-UV irradiation, HCN at large radii requires higher densities and temperatures than the model values. The secondary emission peak (Fig. 1) contributes to this pedestal emission. The far-UV origin of the emission is further supported by the absence of SO emission at the secondary peak, as SO is not much enhanced by far-UV.

Shocks may be an alternative cause for secondary peaks and pedestal emission. The hypothesis is supported by the observed line asymmetry in main isotope HCN. However, the other lines being relatively narrow indicate that most of the structure observed by the interferometer does not originate from shocks. Moreover, single-dish observations do not indicate shocked gas (e.g. van der Tak et al. 2003).

\subsection{Sulfur evaporation}
In addition to protostellar irradiation, the SMA observations also probe sulfur evaporation. The observations indicate that CS and SO line emission extend well into the 900 -- 2500 AU radii region (Figs. 1a and b, Tab. 1). A pronounced abundance enhancement in the range $<$ 3500 AU reduces the ratio of central beam to total SMA flux. Thus, the differences between CS and SO compared to HCN in Tab. 2 are consistent with the CS and SO central peak sizes being more extended due to sulfur evaporation. Such a feature is less pronounced or absent in all three HCN lines. The amount of the abundance jump of CS and SO in the chemical model of St\"auber et al. (2005), however, does not agree when compared with observations in terms of visibilities. The absence of pronounced minima in CS and SO visibilities (Fig. 3) may require a smaller jump than predicted. Instead of reducing the jump magnitude, the effects of the abundance jump may be smoothed by inhomogeneity. If the sulfur evaporation edge varied in radius by $\pm$10\% for different longitudes and latitudes, the superposition of the different visibility distributions would suffice to eliminate the spurious minimum. 

The predicted source sizes of CS and SO are largely determined by the position of the evaporation edge in radius. They exceed the observed sizes of the main peak considerably. Could this be caused by a sulfur evaporation temperature different from the assumed value of 100 K? To qualitatively explore the effect, the chemical model was modified to allow sulfur evaporation at 60 K instead of 100 K. Such a change would double the sulfur evaporation radius and augment the region of enhanced CS and SO abundances. The more distant evaporation edge, however, would make the source even larger than observed and would move the nulls in the model visibilities in Fig. 3a and 3b to $(u,v)$ distances smaller by a factor of two. This would diminish the quality of the fit. The SMA observations suggest a sulfur evaporation radius in the range 700 -- 1300 AU. A higher evaporation temperature is unlikely in view of laboratory experiments (Section 1). Thus, the observations are in agreement with sulfur evaporation near 100 K, roughly coinciding with water evaporation. 

The CS and SO main peak may appear smaller in size because of the above spatial smoothing or a different temperature model. Note that a distance of AFGL 2591 larger than the assumed 1 kpc does not solve the discrepancy, as the angular size of the 100 K radius is independent of distance (S.D. Doty, personal communication). 

\subsection{The innermost envelope}
The agreement of the model with SMA observations of the main peak is acceptable for the unresolved part of the star forming region ($<$ 600 AU in radius). It emits much less than the resolved region, for example less than 5\% of the total main isotope HCN emission. The chemical models overpredict the unresolved HCN($\nu_2$=1) by a factor of three, and underestimate the unresolved part of the main isotope HCN. The observed strong central peak of the HCN conforms with infrared observations by Boonman et al. (2001), indicating an abundance of $10^{-6}$ (Sect. 1). 

We have estimated the optical thickness of the lines from the model density (taken from van der Tak et al. 1999) and abundance (from St\"auber et al. 2005). The column density becomes insensitive to temperature above about 100 K for lines with low upper state energy (thus excluding HCN($\nu_2$=1); van der Tak et al. 2006, Fig. 5). The radiative transfer in molecular lines was calculated applying non-LTE excitation and the escape probability method (RADEX\footnote{http://www.sron.rug.nl/$\sim$vdtak/radex/radex.php}, van der Tak et al. 2007). All the observed lines, even HC$^{15}$N, become optically thick in the center, as noted independently in Fig. 2 for some of the lines. Because of optical thickness and since some of the emission is resolved out by the interferometer, we refrain from estimating molecular column densities based on the SMA observations. 

\section{Conclusions}
Observations of AFGL 2591 at subarcsecond resolution have been made using the SMA in the frequency range 343 -- 354 GHz. The resulting baselines are ideal to explore the envelope of the high-mass YSO at 600 -- 3500 AU radius, a region irradiated by protostellar X-rays and far-UV photons. The observed radii extending over a temperature range of 50 -- 150 K include also the radius of sulfur evaporation from grains. Chemical models predict large changes in abundance of some molecules, such as CS and SO, in this range. The magnitude of the change strongly depends on the flux of possible X-ray irradiation (Fig. 4). The main results are:
\begin{itemize}
\item The visibility amplitudes of SO, the best X-ray tracer according to the model of St\"auber et al. (2005), matches the observations much better when X-rays are included (Fig. 3b). With the exception of HCN($\nu_2$=1) at small radii ($\approx 600$ AU), this holds as well for all other observed molecules. According to the chemical models, the CS abundance is enhanced in the inner envelope and has a significant jump at the sulfur evaporation edge only under the influence of considerable X-ray irradiation (Fig. 4a). Thus the detection of significant extended CS emission in the main peak also corroborates X-ray irradiation. Therefore, these SMA observations support the existence of protostellar X-rays in this HMPO.
\item However, the X-ray model of St\"auber et al. (2005) predicts pronounced minima in visibility amplitude at a $(u,v)$ distance near 110 k$\lambda$ for CS and SO that are not observed. Nevertheless, SO, predicted to have the larger jump, shows a dip at about this baseline that tentatively may be interpreted as a weak signature of it. The model still has to be modified considerably to agree with the CS and SO observations. The jump may be reduced by inhomogeneity in temperature at the radius of sulfur evaporation, varying the radius in a range of at least $\pm 10$\%, so that the minimum is averaged out and the apparent source size increases. Evolutionary effects are alternatives, producing a time delay of grain heating and sulfur evaporation in a non-stationary model. As water evaporates at a similar temperature and radius, a reduction of the jump may then also be expected for H$_2$O.
\item The size of the CS and SO main sources indicate an increase of CS and SO abundance in the range of 100 -- 190 k$\lambda$, corresponding to 1300 -- 700 AU. The results thus concur with sulfur evaporation at 100 K, similar to desorption of water.  
\item The relatively small ratios of central beam SMA flux to total SMA flux (Tab. 2) confirm that CS and SO indeed extend beyond 500 AU. Since the observed FWHM size of the emission is smaller than predicted, this suggests a sulfur evaporation edge at a smaller radius ($\lapprox$ 1000 AU) than modeled. 
\item On the other hand, the visibility amplitudes of CS, HCN, and HC$^{15}$N (but not SO and HCN($\nu_2$=1)) drop in observed visibility from 0 k$\lambda$ (single-dish value) to the lowest $(u,v)$ distance (30 - 40 k$\lambda$) more than predicted (Tab. 2), requiring in addition a broader source component, or pedestal, reaching up to 7000 AU at a low level. 
\item A secondary emission peak located 1180$\pm$140 AU northwest of the main peak within the approaching outflow cone is part of a pedestal emission of the HCN lines from the outflow or its wall. The secondary peak is seen in all HCN lines and also in CS, a molecule that has been found to be sensitive also to far-UV in chemical modeling. The secondary peak is not observed in SO which is insensitive to far-UV irradiation. The pedestal emission is clearly indicated also in the visibility amplitudes of CS, HCN, and HC$^{15}$N,  but not SO and HCN($\nu_2$=1).
\item The line emission from the outflow or its walls is in all cases less than 30\% of the total interferometric flux.
\item The visibility analysis of the SMA data indicates that CS emission has a more pronounced unresolved central peak than SO. This agrees with the predictions of the chemical models (Figs. 4a and b). The observed unresolved emission is slightly larger than modeled for CS and the main isotope HCN. 
\end{itemize}

In conclusion, we present the first quantitative spatially resolved tests of an X-ray model for YSO envelopes previously fitted to unresolved single-dish data (St\"auber et al. 2005). The SMA observations support the hypothesis of
protostellar X-rays and far-UV irradiation, but require an additional broad pedestal component possibly located in the outflows. Overall, these observations demonstrate the power of submillimeter interferometers for
directly testing abundance profiles of molecules in protostellar
envelopes on scales where many important physical and chemical processes are
happening.

For a quantitative description of the radial abundance variations in the envelope it will be necessary to fit a detailed, possibly non-spherically symmetric model to imaged data. This would require a large computational effort, beyond the scope of this work, but could potentially provide a more detailed picture for the influence of X-ray and far-UV protostellar emission compared to other chemical effects in high-mass protostars.

\begin{acknowledgements}
We acknowledge helpful discussions with S.D. Doty and thank also the SMA staff for the continued development of this instrument and, in particular Alison Peck, for invaluable help with the observations. The submillimeter work at ETH is supported by the Swiss National Science Foundation grant 200020-105366.
\end{acknowledgements}


{}

\end{document}